\documentclass[supplement]{ptptex}

\usepackage{graphicx}
\title{Current and vorticity auto correlation functions in open microwave billiards}
\author{
Y.-H.~\textsc{Kim}, M.~\textsc{Barth}, U.~\textsc{Kuhl}, and H.-J.~\textsc{St\"ockmann}
\footnote{e--mails:
young-hee.kim@physik.uni-marburg.de, michael.barth@physik.uni-marburg.de,\\
ulrich.kuhl@physik.uni-marburg.de, stoeckmann@physik.uni-marburg.de } } \inst{Fachbereich
Physik der Philipps-Universit\"at Marburg, Renthof 5,
\newline
D-35032 Marburg, Germany}

\abst{Using the equivalence between the quantum-mechanical probability density in a
quantum billiard and the Poynting vector in the corresponding microwave system, current
distributions were studied in a quantum dot like cavity, as well as in a Robnik billiard
with $\lambda=0.4$, and an introduced ferrite cylinder. Spatial auto correlation
functions for currents and vorticity were studied and compared with predictions from the
random-superposition-of-plane-waves hypothesis. In addition different types of vortex
neighbour spacing distributions were determined and compared with theory.}

\begin{document}

\maketitle

\section{Introduction}
\label{sec:Introduction}

The large majority of wave functions of chaotic billiards is chaotic, i.\,e.\ at any
point in the system, not too far from the wall, the wave function may be well described
by a random-superposition of plane-waves (RSPW) \cite{Ber77a},

\begin{equation}
\label{eq:rspw} \psi(\vec{r})=\sum_n a_n\exp{^{i\vec{k}_n\vec{r}}},
\end{equation}
where modulus $k=|\vec{k}_n|$ of the incoming wave is fixed, but directions $\vec{k}_n/k$
and amplitudes are considered as random. As an immediate consequence the wave function
amplitudes are Gaussian distributed, or, equivalently, their squares $\rho=|\psi|^2$ are
Porter-Thomas distributed,

\begin{equation}
\label{eq:PT} P(\rho)=\sqrt{\frac{A}{2\pi \rho}}  \exp{(-\frac{A}{2}\rho)},
\end{equation}
where $A$ is the billiard area. For the spatial correlation function of the wave function
amplitudes one obtains a Bessel function,

\begin{equation}
\label{eq:CPsi2} C(\vec{r}_1,\vec{r}_2) = \frac{
\left\langle\psi^*(\vec{r}_1)\psi(\vec{r}_2)\right\rangle }{
|\left\langle\psi(r)\right\rangle|^2 } = J_0(kr),
\end{equation}
where $r=|\vec{r}_1-\vec{r}_2|$. The brackets denote an average over all positions. All
these features have been demonstrated by McDonald and Kaufman in their influential work
on stadium wave functions \cite{McD79,McD88}. Very recently Urbina and Richter
\cite{Urb} succeeded in giving a quantum mechanical justification of the approach,
generalizing ideas of Hortikar, Srednicki \cite{Hor98a}, and Gornyi, Mirlin \cite{Gor02b}. It
is impossible to mention all papers which have been published hitherto on the subject.
The RSPW approach is not restricted to quantum mechanics. This is why experiments using
classical waves have played an important role, since for a long time they were the only
ones with the ability to look into the system. Very recently techniques have been
developed which yield comparable information for electron flow patterns in mesoscopic
structures \cite{Top00}. The state of the art of the experiments with classical waves up
to the year 1999 is presented in reference \citen{Stoe99}.

Most of the experiments with classical waves have been performed in microwave resonators
\cite{Stoe90,Sri91,Grae92a} and vibrating solids \cite{Ell95}. In one work light
propagation through a wave guide with distorted cross-section was studied \cite{Doy02a}.
In all cases the predictions of the RSPW approach could be verified. It should be noted
that in the general case there is {\it no} one-to-one correspondence to quantum
mechanics, thus demonstrating the universality of the approach. Therefor similar ideas
have been developed independently in the context of room acoustics \cite{Ebe84}.
Quasi-two-dimensional microwave resonators constitute one prominent exception where the
equivalence to quantum mechanics is complete, including the boundary conditions. This is
no longer true in three-dimensional resonators. But even here the approach remains valid
\cite{Doer98b}. One only has to superimpose plane {\it electromagnetic} waves with the
consequence that expression (\ref{eq:CPsi2}) for the spatial autocorrelation function has
to be modified \cite{Eck99}. Of course there are limitations. Since the RSPW approach
completely ignores boundary conditions, deviations are expected and found in regions
close to the boundary \cite{Baec02d,Ber02b,Baec02a}, or if the wavelengths are not small
compared to the system size.

The approach definitely cannot be applied to wave functions which are scarred along
periodic orbits or show regular patterns associated with bouncing balls \cite{Hel84}. It
was shown already by McDonald and Kaufman that for such wave functions the wave function
amplitudes are {\it not} Gaussian distributed \cite{McD88}. If the billiards are open, or
if time-reversal symmetry is broken, the wave function are complex, and currents are
present. In microwave experiments wall absorption is another source of currents. The
quantum-mechanical probability density is given by

\begin{equation}
\label{eq:4} j(r)=\mathrm{Im}(\psi^*\nabla \psi).
\end{equation}
In quasi-two-dimensional electromagnetic cavities there is a one-to-one correspondence to
the Poynting vector making an experimental determination of $j(r)$ feasible as well
\cite{Seb99}.

The consequences of the RSPW approach for the distribution of currents have been studied
in particular by Berggren and coworker in a series of papers
\cite{Ber99a,Sai01,Ber01b,Sai02b}. In open systems there are no longer nodal lines but
nodal points, or vortices, since for the wave function to be zero both real and imaginary
part have to be zero at the same time. Two-point correlation functions of vortices have
been given independently by Berry and Dennis \cite{Ber00b} and by Saichev et
al.~\cite{Sai01}. Nearest neighbour distributions of vortices have been studied in
Ref.~\citen{Ber02a}. The theoretical predictions have been tested experimentally in two
microwave experiments \cite{Bar02,Vra02}, including a direct visualization of persistent
currents well-known from mesoscopic physics. In the present paper a number of additional
microwave tests of the RSPW hypothesis are presented with special emphasis on spatial
auto correlation functions of currents and vorticities. To the best of our knowledge such
quantities have never been studied before, neither theoretically nor experimentally.
After a short recapitulation of the experimental technique in
section~\ref{sec:Experiment}, analytic expressions for a number of autocorrelation
function are derived in section~\ref{sec:Theory}. Various comparisons between experiment
and theory are presented in section~\ref{sec:Results}.

\section{Experiment}
\label{sec:Experiment}

\begin{figure}
\includegraphics[width=14cm]{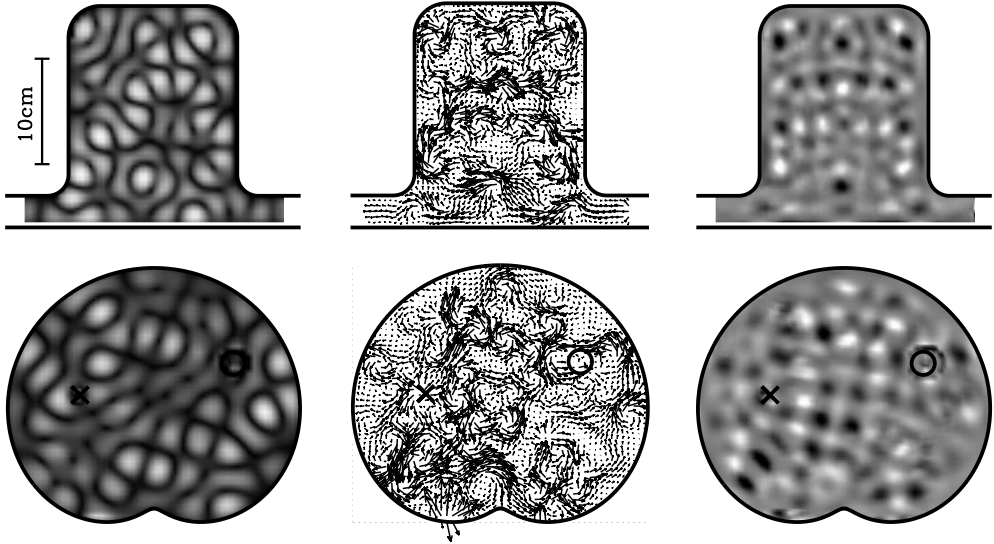}
\caption{\label{fig:PsiJVort} Plot of wave function amplitude square $|\Psi|^2$,
probability density $\vec{j} = \mathrm{Im}(\Psi^*\nabla\Psi)$, and vorticity
$\omega=(\nabla_x \Psi_R)(\nabla_y \Psi_i)-(\nabla_y \Psi_R)(\nabla_x \Psi_i)$ for the
quantum dot billiard (upper row) at $\nu$=~8.33~GHz, and the Robnik billiard with a
ferrite ring insert (lower row) at $\nu$=~7.03~GHz. The position of the antenna is
indicated by a cross ($\times$).
% and a circle($\circ$).
\newline
In the $|\Psi|^2$ plot (left column) the intensity is converted into a gray scale, where
black corresponds to zero intensity. In the vorticity plot (right column) black
corresponds to a large positive, and white to a large negative sense of rotation. }
\end{figure}

In the description of the experiment we can be short, since all details are available
from our previous publications for the Robnik \cite{Bar02,Vra02}, and the quantum dot
billiard \cite{Kim02}. The quantum dot billiard is an open billiard of rectangular shape
with rounded corners, and an entrance and an exit wave guide attached at opposite sides
(see Fig.~\ref{fig:PsiJVort}). The second one belongs to the family of the Pascal
lima\c{c}on, or Robnik billiards. It can be obtained by a complex mapping of the unit
circle by means of the function $\omega=z+\lambda z^2$. For the parameter $\lambda=0.4$,
used in the experiment, the classical phase space is completely chaotic \cite{Rob83},
apart from, perhaps, tiny regular fractions \cite{Dul01}. A ferrite ring has originally
been introduced to break time reversal symmetry thus giving rise to persistent currents.
Time reversal symmetry is broken only for a small frequency region (approximately 3 to
5~GHz) but the absorption of the ferrite is present for the whole frequency range used
for the data analysis (3 to 10~GHz). In the present context it is not of relevance
whether the origin of the currents is break of time-reversal symmetry or absorption.

The linear dimensions of both cavities is of the order of 25~cm (see the scale in
Fig.~\ref{fig:PsiJVort}), and their height is $h$=8~mm. For frequencies below 18.75~GHz
there is a complete equivalence between the quantum mechanical wave function amplitudes
$\psi$ and the electric field $E_z$, where the quantum-mechanical eigenenergy $E$
corresponds to the square of the wavenumber $k^2$. Measuring the reflection amplitude at
one antenna, or the transmission amplitude between two antennas, the complete scattering
matrix can be obtained, which for isolated resonances reduces to a billiard Breit-Wigner
function \cite{Ste95}

\begin{equation}
S_{ij}=\delta_{ij}-2i\gamma \sum_n
\frac{\mathrm{Im}[\psi_n^*(r_i)\psi_n(r_j)]}{k^2-k_n^2} \label{eq:5}
\end{equation}

Both eigenfunctions and eigenenergies are slightly modified by the presence of the
antenna which has been neglected in Eq.~(\ref{eq:5}) (for an introductory presentation
see chapter 6 of Ref.~\citen{Stoe99}). From a transmission measurement thus the wave
function can be obtained including the sign, whereas a reflection measurement only yields
the modulus.

For quasi-two-dimensional systems the Poynting vector
$\vec{S}=c/(4\pi)\vec{E}\times\vec{H}$ reduces to \cite{Seb99}

\begin{equation}
\label{eq:6} \vec{S}=\frac{c}{8\pi k} \mathrm{Im}[E_z^*(r)\vec{\nabla}E_z(r)],
\end{equation}
illustrating the one-to-one correspondence to the quantum mechanical probability density
(4) stated above. Typical results for the two billiard systems under study are shown in
Fig.~\ref{fig:PsiJVort}. Further examples can be found in our previous publications
\cite{Bar02,Vra02}. In the right column $|\psi|^2$ is plotted in a gray scale for a
typical frequency. The middle column shows the corresponding flow pattern as obtained
from Eq.~(\ref{eq:6}), where the arrows reflect the Poynting vectors at the respective
points. The right column shows a plot of the vorticities, or vortex strengths. The
vorticity is, up to the factor 1/2, just the curl of the current \cite{Ber00b} and
reduces for two-dimensional systems to

\begin{equation}
\label{eq:7} \omega=(\nabla_x\psi_R)(\nabla_y\psi_I) -
    (\nabla_y\psi_R)(\nabla_x\psi_I),
\end{equation}
where $\psi_R,\psi_I$ are real and imaginary part of the wave function. A plot of the
vorticity is particularly useful to make the vortex pattern visible as is evident from
Fig.~\ref{fig:PsiJVort}.

\section{Theory}
\label{sec:Theory}

It follows immediately from the RSPW hypothesis, as a consequence of the central limit
theorem, that the $\psi(r)$ can be treated as Gaussian random variables. They obey the
well-known property that all higher moments can be expressed in terms of the second
moment. Thus all distributions of interest can be calculated \cite{Sre96a,Sre96b}. We do
not follow this route, however, mainly for pedagogical reasons, but start directly from
Eq.~(\ref{eq:rspw}) to calculate current and vorticity auto correlation functions.

Writing $\vec{k}_n=k(\cos\varphi_n,\sin\varphi_n)$, we obtain for the derivatives of the
wave function

\begin{eqnarray}
\label{eq:8}
\frac{\partial\psi}{\partial x}&=& ik \sum_n a_n \cos\varphi_n~e^{i\vec{k}_n \vec{r}}\nonumber\\
\frac{\partial\psi}{\partial y}&=& ik \sum_n a_n \sin\varphi_n~e^{i\vec{k}_n \vec{r}}
\end{eqnarray}

Using Eq.~(\ref{eq:4}), it follows for the $x$ component of the current

\begin{equation}
\label{eq:9} j_x(\vec{r})=k\sum_{n,m}a_n^*a_m(\cos\varphi_n+\cos\varphi_m)
                e^{-i(\vec{k}_n-\vec{k}_m)\vec{r}}.
\end{equation}

In calculating the autocorrelation functions

\begin{equation}
\label{eq:10} C_{j_x}(\vec{r_1},\vec{r_2})\sim \left\langle j_x(\vec{r}_1)j_x(\vec{r}_2)
\right\rangle
\end{equation}
we use the assumption that the $a_n$ are uncorrelated
\begin{equation}
\label{eq:11} \left\langle a_n^*a_m \right\rangle=\left\langle |a_n|^2 \right\rangle
\delta_{nm}.
\end{equation}
It follows from Eqs.~(\ref{eq:9}) and (\ref{eq:10})

\begin{eqnarray}
\label{eq:12} \hspace*{-1cm} C_{j_x}(\vec{r_1},\vec{r_2})&\sim & k^2
    \left\langle \sum_{n,m}|a_n|^2|a_m|^2(\cos\varphi_n+\cos\varphi_m)^2
                e^{-i(\vec{k}_n-\vec{k}_m)(\vec{r}_1-\vec{r}_2)}\right\rangle
\nonumber\\
        &\sim & \left\langle \cos^2\varphi_n e^{-i\vec{k}_n\vec{r}}\right\rangle
                \left\langle e^{i\vec{k}_n\vec{r}}\right\rangle+
                \left\langle\cos\varphi_n e^{-i\vec{k}_n\vec{r}}\right\rangle
                \left\langle\cos\varphi_n e^{i\vec{k}_n\vec{r}}\right\rangle,
\end{eqnarray}
where $\vec{r}=\vec{r}_1-\vec{r}_2$. All averages can be expressed in terms of Bessel
functions with the result

\begin{equation}
\label{eq:13} C_{j_x}(r)=\left\langle J_0(kr)[J_0(kr)-\cos 2\varphi J_2(kr)]+2\cos^2
\varphi[J_1(kr)]^2\right\rangle,
\end{equation}
where we have written $C_{j_x}(r)$ instead of $C_{j_x}(\vec{r_1},\vec{r_2})$ to indicate
that the autocorrelation function depends on $r=|\vec{r}_1-\vec{r}_2|$ exclusively. The
normalization $C_{j_x}(0)=1$ was applied. It only remains to perform the average over
$\varphi$, the angle between vector $\vec{r}$ and the $x$ axis. The averaging gives

\begin{equation}
\label{eq:14} C_{j_x}(r)=[J_0(kr)]^2+[J_1(kr)]^2.
\end{equation}

For the autocorrelation function of $j_y(r)$ the same expression is obtained. Instead of
averaging over $\varphi$, we may alternatively look for two other quantities, namely the
autocorrelation functions of $j_{\parallel}(r)$, and $j_{\bot}(r)$, the current
components parallel and perpendicular to $\vec{r}$, respectively. Inserting $\varphi=0$
and $\varphi=\frac{\pi}{2}$ we obtain from Eq.~(\ref{eq:13})

\begin{eqnarray}
\label{eq:15} C_{j_{\parallel}}(r) &=& J_0(kr)[J_0(kr)-J_2(kr)]+2J_1(kr)
\nonumber\\
C_{j_{\bot}}(r)      &=& J_0(kr)[J_0(kr)+J_2(kr)]
\end{eqnarray}

In the same way the spatial autocorrelation function for the vorticity is obtained,
entering definition (7) with expressions (8) for the derivatives of the wave function.
The derivation is a step-by-step repetition of the calculation for $C_{j_x}(r)$, and we
find

\begin{eqnarray}
\label{eq:16} C_\omega(\vec{r}) &=&
\frac{\left\langle\omega(\vec{r}_1)\omega(\vec{r}_2)\right\rangle}{\left\langle
\omega(\vec{r}_1)^2\right\rangle}
\nonumber\\
                &=&[J_0(kr)]^2-[J_1(kr)]^2
\end{eqnarray}

\section{Results}
\label{sec:Results}

It was explained in Ref.~\citen{Ber02a} that there is a problem with the experimental
determination of field and current distributions: the probe antenna moving though the
billiard gives rise to a leakage current, which is critical close to positions where the
field amplitudes are large. For the quantum dot billiard in addition there are
frequencies where the total transmission is zero. In such a case the situation is even
worse, since now there is exclusively the leakage current from the entrance to the probe
antenna. All these problematic frequency regions have been omitted from the data
analysis.

\begin{figure}
\includegraphics[width=14cm]{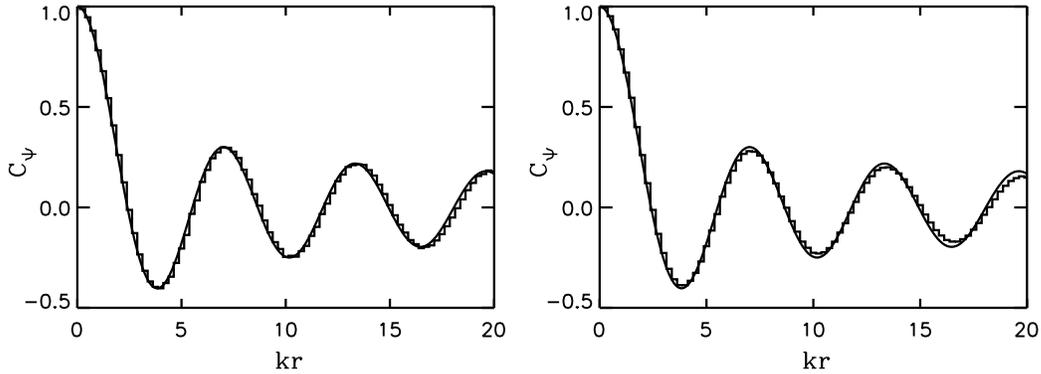}
\caption{\label{fig:C_Psi2} Experimental spatial autocorrelation functions of the wave
function amplitude for the quantum dot billiard (left) and the Robnik billiard with
ferrite insert (right). The solid lines correspond to the prediction from the
random-superposition of plane-waves approach (see Eq.~(\ref{eq:CPsi2})). }
\end{figure}

For the quantum dot billiard there is another problem. In the low-frequency regime the
wave functions are still reminiscent of the rectangle with the checkerboard patterns
typical for such systems. In addition there are frequencies showing strong scarring
associated with bouncing balls and classical trajectories through the system. (The
relation between scarring and transport was our original motivation to study this system
\cite{Kim02}, triggered by observations in correspondingly shaped quantum dots). This is
why for the quantum dot billiard only frequencies above 4.2~GHz were considered, where
the system behaves chaotically.

\begin{figure}
\includegraphics[width=14cm]{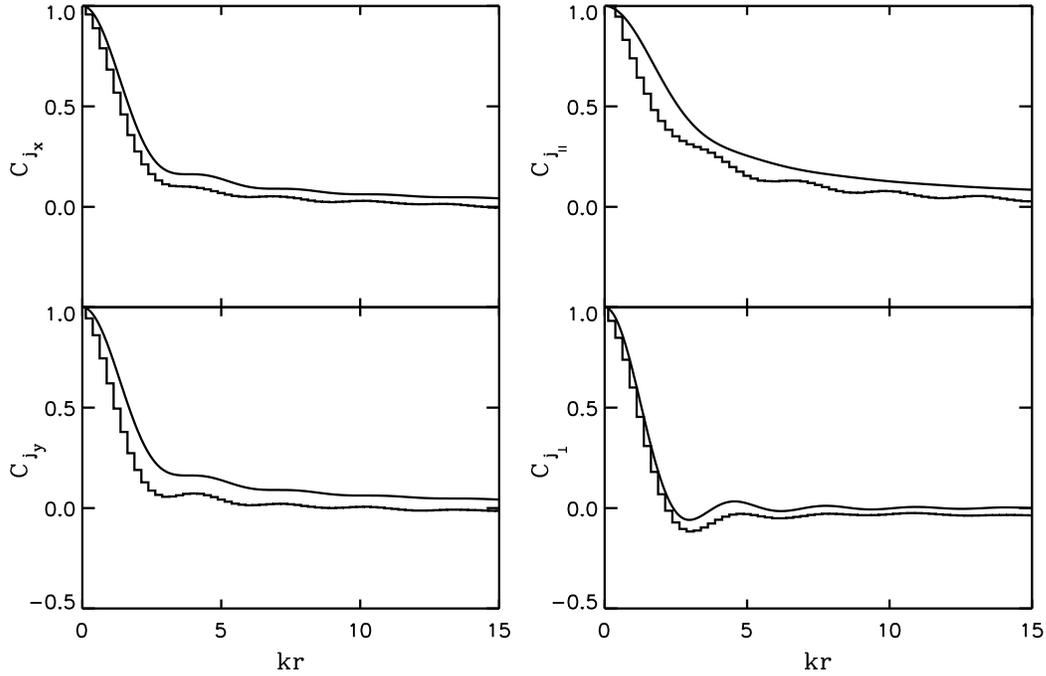}
\vspace*{-0.5cm} \caption{\label{fig:Cjxjyjpjv} Experimental spatial autocorrelation
functions of $j_x(r)$, $j_y(r)$, $j_\parallel(r)$, and $j_\bot(r)$ (from top to bottom)
for the quantum dot billiard (left column) and the Robnik billiard with ferrite insert
(right column). The solid lines correspond to the predictions from the
random-superposition of plane-waves approach (see Eqs.~(\ref{eq:14}) to (\ref{eq:15})). }
\end{figure}

\begin{figure}
\includegraphics[width=14cm]{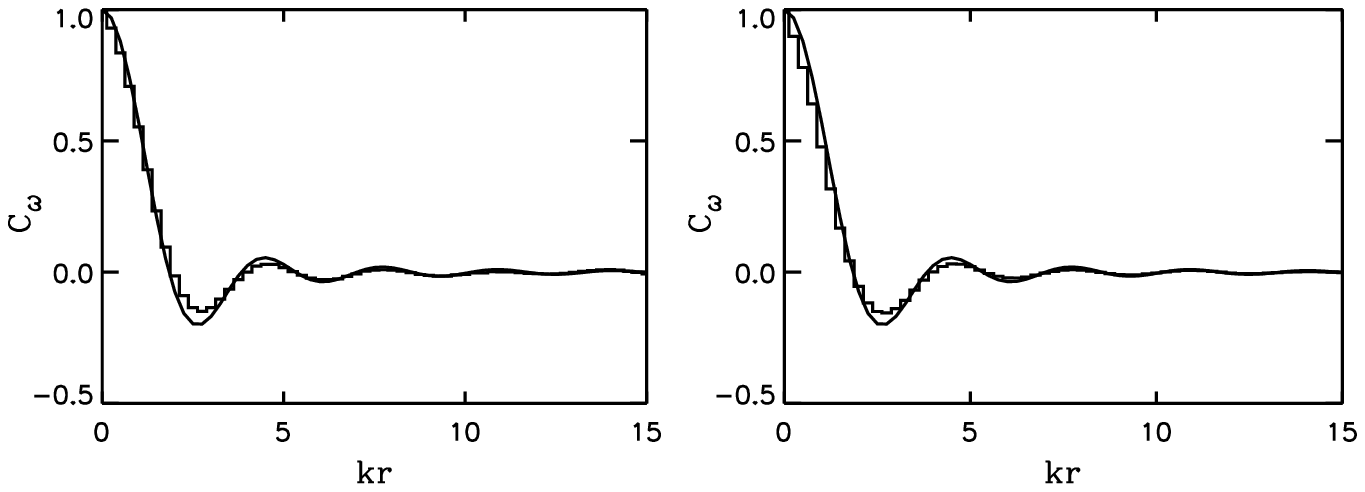}
\caption{\label{fig:vort_ac} Experimental spatial autocorrelation functions of the
vorticity $\omega(r)$ for the quantum dot billiard (left) and the Robnik billiard with
ferrite insert (right). The solid lines correspond to the predictions from the
random-superposition of plane-waves approach (see Eq.~(\ref{eq:16})). }
\end{figure}

All correlation functions discussed in section~\ref{sec:Theory} depend on the parameter
$kr$ exclusively. Therefore it is possible to superimpose the results from different
frequencies by an appropriate rescaling to improve statistics. We start with the
presentation of our results for the spatial autocorrelation function of the wave function
amplitudes (see Fig.~\ref{fig:C_Psi2}). A perfect agreement is found for both systems
between the experimental results and the prediction from the RSPW hypothesis. This may be
considered as a check for the validity of the approach in the selected frequency regimes.

In Fig.~\ref{fig:Cjxjyjpjv}, the results of the different current autocorrelation
functions introduced in section \ref{sec:Theory}, are shown for the Robnik billiard with
ferrite insert. The corresponding figures for the quantum dot billiard have been omitted,
since the results for the systems are more or less identical. Though there are deviations
in detail, the overall qualitative agreement is very good. In particular the
qualitatively different behaviour for the various types of current autocorrelation
functions is reproduced correctly. For the vorticity autocorrelation function, shown in
Fig.~\ref{fig:vort_ac}, the agreement between theory and experiment is nearly perfect.
One only can speculate why this is the case: To determine the current, one needs the
product of the wave function with its derivative, whereas the vorticity is obtained from
the product of two derivatives. As a consequence, in the latter case all current offsets
slowly varying with the position are eliminated. Therefore the vorticity is less
sensitive on the mentioned experimental imperfections, as it seems.

\begin{figure}
\includegraphics[width=14cm]{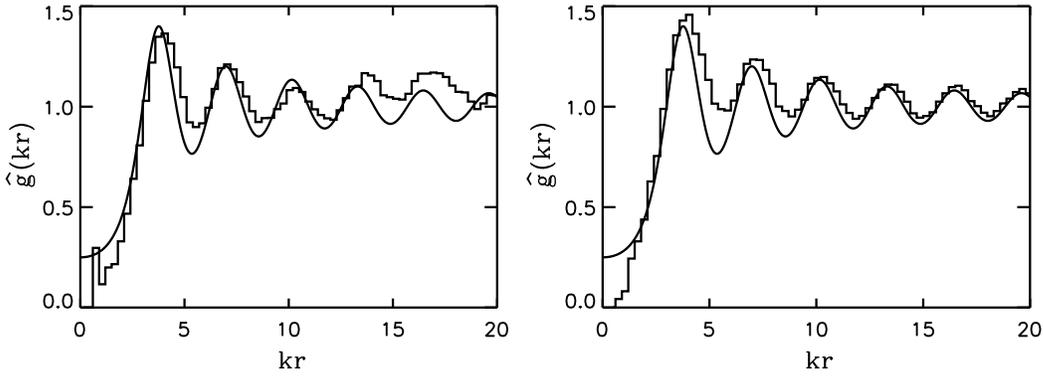}
\caption{\label{fig:CVort} Vortex pair correlation function for the quantum dot billiard
(left) and the Robnik billiard with ferrite insert (right). The solid line corresponds to
the theoretical prediction from the random-superposition-of-plane waves approach
\cite{Ber00b,Sai01}. }
\end{figure}

In Fig.~\ref{fig:CVort} vortex pair correlation functions for both systems are presented.
Preliminary results for the Robnik billiard have already been shown in
Ref.~\citen{Ber02a}, where a more detailed discussion of this quantity can be found. The
expected oscillatory behaviour again is reproduced correctly. The deviations between
experiment and theory at small distances reflect the experimental resolution. The data
were taken on a grid of 5 mm side length with the consequence that vortices with a
distance below 10 mm can no longer be separated reliably.

\begin{figure}
\includegraphics[width=14cm]{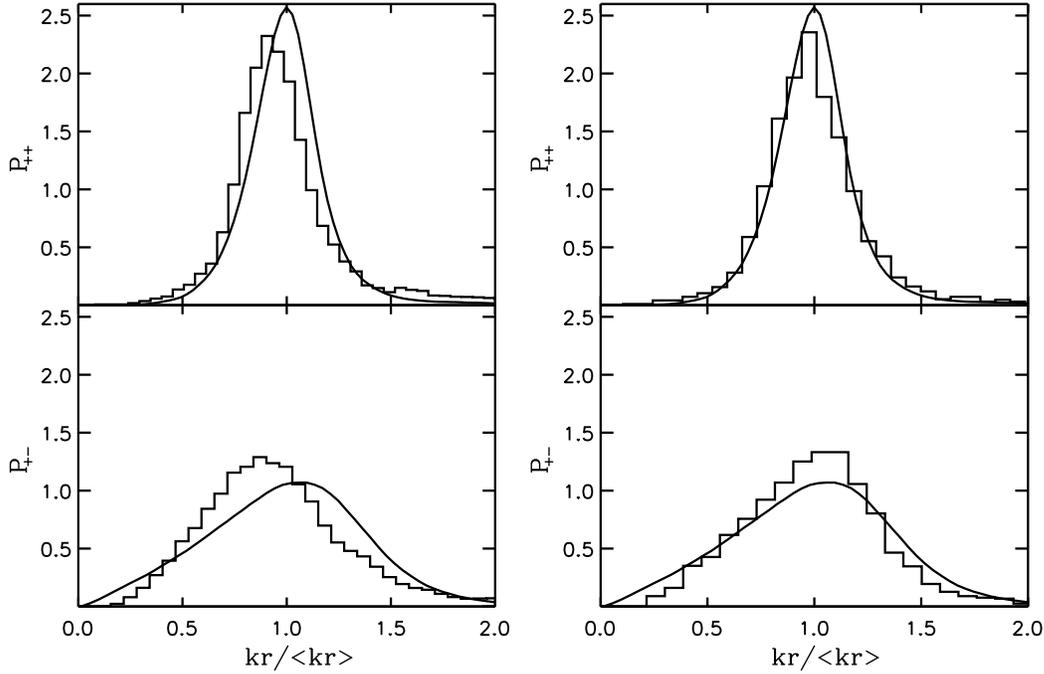}
\vspace*{-0.5cm} \caption{\label{fig:P++P+-} Nearest neighbour distance distribution of
vortices of opposite ($P_{+-}(r)$), and same sense of rotation ($P_{++}(r)$),
respectively, for the quantum dot billiard (left), and the Robnik billiard with ferrite
insert (right). The solid lines corresponds to the numerical evaluation of the nodes in
the Berry function. }
\end{figure}

We now extend this discussion to the nearest neighbour spacing distribution between
vortices. This quantity was introduced by Saichev et al. \cite{Sai01} and was studied by
the authors in a number of papers. There are different types of spacing distributions
denoted by $P_{++}(r), P_{+-}(r),P_{-+}(r),P_{--}(r)$ where the pair of indices denotes
the sense of rotation of the vortices considered. $P_{++}(r)$ and $P_{--}(r)$, as well as
$P_{+-}(r)$ and $P_{-+}(r)$ should be identical, of course. Only for small system sizes
(where the RSPW approach fails anyway) there may be deviations due to the presence of the
boundary \cite{Sai01}. Fig.~\ref{fig:P++P+-} shows our results, for $P_{++}(r)$ and
$P_{+-}(r)$. Within the limits of statistical errors there was no difference to the
corresponding distributions of $P_{--}(r)$ and $P_{-+}(r)$, respectively. The theoretical
curve is the result of direct numerical evaluation of the nodes calculated via RSPW
approach (see \cite{Ber77a,Sai01}). The deviations between experiment and theory for
$P_{+-}(r)$ are comparable to that found in the papers of the Berggren group
\cite{Sai01,Ber02a}, and reflect at least partly the limitations of the Poisson
approximation. Another cause, in particular for the small distances, is the limited
experimental resolution discussed above. For $P_{++}(r)$, on the other hand, the
agreement between experiment and theory is good, again in accordance with
Refs.~\citen{Sai01,Ber02a}. In addition, the experiment resolution has only a small
effect in this case, since small distances do not contribute anyway to $P_{++}(r)$
significantly.

\section{Conclusion}
\label{sec:Conclusion}

In this paper a number of consequences of the RSPW hypothesis have been presented not
been studied hitherto. A qualitatively good agreement between the experiment and the
theoretical prediction was found for different types of spatial current correlation
functions. The remaining discrepancies are probably due to experimental imperfections
caused by leakage currents into the probe antennas. For the vorticity autocorrelation
function a perfect agreement between experiment and theory is found. In addition vortex
pair correlation functions, and vortex nearest neighbour distance distributions were
studied. Again the agreement between experiment and theory are good, apart from
discrepancies at small distances.

\section*{Acknowledgements}

The experiments were supported by the Deutsche Forschungsgemeinschaft. Rudi Sch\"afer,
Marburg is thanked for helpful discussions. Part of the results presented in this paper
were presented on the 5th International Summer school 2002 in Maribor. M. Robnik and his
group is thanked for hospitality. The participation in the summer school was supported by
the scientific and academic cooperation program between the universities of the twin
towns Maribor and Marburg. K.-F. Berggren, Link\"oping is thanked for making his
calculations of the vortex distance distributions available to us, for critical
reading, and for calling our attention to reference \citen{Ebe84}.

\bibliographystyle{prsty_long}
\bibliography{thesis,paperdef,paper,newpaper,book}

\end{document}